\documentclass[aps,showpacs,pra,superscriptaddress,twocolumn,floatfix,10pt,longbibliography]{revtex4-1}

\usepackage[colorlinks]{hyperref}
\hypersetup{citecolor=blue,linkcolor=blue}

\usepackage[a4paper, left=.7in, right=.7in, top=.7in, bottom=.9in]{geometry} 

\usepackage{tabularx}
\usepackage{ctable} % for \specialrule command

\usepackage{amsfonts}
\usepackage{amsmath}
\usepackage{amssymb}
\usepackage{times}
\usepackage{hhline}

\newcommand{\x}{\bs{x}}

\newcommand{\J}{\bs{J}}
\newcommand{\bs}[1]{{\boldsymbol{#1}}}
\newcommand{\PsiI}{\Psi^\text{i}}
\newcommand{\PsiR}{\Psi^\text{r}}

\newcommand{\phiI}{\phi^\text{i}}
\newcommand{\phiR}{\phi^\text{r}}
\newcommand{\T}{\mathcal{T}}
\newcommand{\F}{\mathcal{F}}
\renewcommand{\L}{\mathcal{L}}
\renewcommand{\H}{{H}}
\renewcommand{\P}{\mathcal{P}}

\begin{document}

\title{Generalized continuity equations from two-field Schr\"odinger Lagrangians}

\author{A.~G.~B.~Spourdalakis}
\affiliation{Department of Physics, University of Athens, 15771 Athens, Greece}

\author{G.~Pappas}
\affiliation{Department of Physics, University of Athens, 15771 Athens, Greece}

\author{C.\,V.~Morfonios}
\affiliation{Zentrum f\"{u}r Optische Quantentechnologien, Universit\"{a}t Hamburg, 22761 Hamburg, Germany}

\author{P.~A.~Kalozoumis}
\affiliation{Department of Physics, University of Athens, 15771 Athens, Greece}

\author{F.~K.~Diakonos}
%\email[]{fdiakono@phys.uoa.gr}
\affiliation{Department of Physics, University of Athens, 15771 Athens, Greece}

\author{P.~Schmelcher}
%\email[]{pschmelc@physnet.uni-hamburg.de}
\affiliation{Zentrum f\"{u}r Optische Quantentechnologien, Universit\"{a}t Hamburg, 22761 Hamburg, Germany}
\affiliation{The Hamburg Centre for Ultrafast Imaging, Universit\"{a}t Hamburg, 22761 Hamburg, Germany}
\date{\today}

\begin{abstract}
A variational scheme for the derivation of generalized, symmetry-induced continuity equations for Hermitian and non-Hermitian quantum mechanical systems is developed. 
We introduce a Lagrangian which involves two complex wave fields and whose global invariance under dilation and phase variations leads to a mixed continuity equation for the two fields. 
In combination with discrete spatial symmetries of the underlying Hamiltonian, the mixed continuity equation is shown to produce bilocal conservation laws for a single field.
This leads to generalized conserved charges for vanishing boundary currents, and to divergenceless bilocal currents for stationary states.
The formalism reproduces the bilocal continuity equation obtained in the special case of $\P\T$ symmetric quantum mechanics and paraxial optics. 
\end{abstract}

%\pacs{}
\maketitle

\section{Introduction}\label{sec:1}

The variational principle of stationary action provides a generic path connecting a representation of a physical system by a scalar function, the Lagrangian, with the equations of motion (EOM) determining its time evolution. 
Although this relation is not bijective (different Lagrangians may lead to the same EOM), all Lagrangians representing a given physical system have a common property: their actions remain invariant under the same transformations of the involved variables.
Symmetries of the Lagrangian under \textit{continuous} transformations then manifestly lead to corresponding conservation laws via Noether's theorem \cite{Noether1918_NVGWZ_1918_235_InvarianteVariationsprobleme},
which can be seen as operating in a twofold way: It (i) provides the form of quantities obeying symmetry-induced conservation laws and, conversely, (ii) dictates the symmetric design of the Lagrangian of a system under predefined conservation laws. 
A typical example of the latter case is the $U(1)$ symmetric Lagrangian constructed to provide a variational formulation of quantum mechanics, leading to the continuity equation for the probability current.

The symmetry of a system under \textit{discrete} spatial transformations also yields conserved quantities, though now by commutation of the corresponding operator with the Hamiltonian rather than through Lagrangian variation.
In the context of non-Hermitian quantum systems with symmetry under combined spatial ($\P$) and temporal ($\T$) reflection, a symmetry-induced nonlocal continuity equation can be derived from the Schr\"odinger equation (SE) \cite{Bagchi2001_MPLA_16_2047_GeneralizedContinuity}.
This alternative description further necessitates the introduction of a suitably defined nonlocal scalar product reflecting the $\P\T$ transformation in consistently obtained expectation values \cite{Japaridze2002_JPAMG_35_1709_SpaceState}.
$\P\T$-induced conservation laws were also obtained recently in the context of nonlinear systems with self-induced $\P\T$ symmetry by applying Noether's theorem to the associated nonlocal Lagrangian possessing a set of continuous symmetries \cite{Sinha2015_PRE_91_042908_SymmetriesExact}, with the $U(1)$ symmetry also treated in paraxial optics \cite{El-ganainy2007_OL_32_2632_TheoryCoupled}.
$\P\T$ symmetry and its spontaneous breaking \cite{Chong2011_PRL_106_093902_Mathcalpmathcalt-symBreaking} has received increased attention since its realization in photonic heterostructures, featuring phenomena such as anisotropic transmission resonances~\cite{Ge2012_PRA_85_023802_ConservationRelations}, coherent perfect absorber laser points~\cite{Longhi2010_PRA_82_031801_Mathcalpt-symmetricLaser}, unidirectional invisibility~\cite{Lin2011_PRL_106__UnidirectionalInvisibility}, or absorption enhanced transmission~\cite{Guo2009_PRL_103__ObservationP}, to mention a few. 
The $\P\T$-adapted nonlocal current of stationary states can here be employed as a natural order parameter \cite{Kalozoumis2014_PRA_90_043809_SystematicPathway,Kalozoumis2016_PRA_93_063831_Mathcalpt-symmetryBreaking} for the symmetry-breaking transition.

Nonlocal conservation laws are, nevertheless, not exclusively linked to $\P\T$ symmetric systems.
Indeed, Ref.\,\cite{Kalozoumis2014_PRL_113_050403_InvariantsBroken} uses two \textit{different} nonlocal currents to generalize the amplitude mapping in parity and Bloch eigenstates from global to local symmetries in finite domains.
The spatial constancy of those stationary currents, derived there from the stationary SE, suggests their origin in suitably generalized conservation laws applied to inversion and translation symmetry, respectively.
In view of the above, a natural question which arises is whether such generalized conservation laws adapted to discrete symmetries may arise from a common variational principle.

In the present work we answer this question by introducing a two-field Lagrangian whose invariance under global dilatation and phase transformation leads to a mixed continuity equation for the two fields. 
The treatment is inspired by the method of phase space extension for dissipative systems \cite{Bateman1931_PR_38_815_DissipativeSystems}, with the two states correlated in the Lagrangian here being generally solutions to dual Schr\"odinger equations with opposite imaginary potential terms modeling loss and gain.
Together with the symmetry of the Hamiltonian under a discrete spatial transformation, potentially combined with time reversal, the mixed continuity equation produces corresponding generalized nonlocal current conservation laws for a single state.
Those apply to any wave mechanical system described effectively by a SE, such as optics in two or three dimensions within the paraxial approximation.
In particular, the concept of optical quasipower, used in the literature in the context of $\P\T$ symmetry, is here generalized to, e.\,g., purely lossy systems with arbitrary discrete symmetries. 
In one dimension, the spatially constant nonlocal currents of Ref.\,\cite{Kalozoumis2014_PRL_113_050403_InvariantsBroken} for stationary states are recovered. 
The formalism addresses Hermitian and non-Hermitian systems on equal footing in arbitrary dimensions, and its extension to symmetric interacting Hamiltonians is straightforward.
It thus provides a unified theoretical framework for the variational extraction of conservation laws for wave systems with discrete spatial symmetries.

The paper is organized as follows. 
In Sec.~\ref{sec:correlator_lagrangian} we introduce the two-field Lagrangian leading to two complementary SEs with opposite imaginary potential via the variational principle.
In Sec.~\ref{sec:mixed_cont_eq} Noether's theorem is applied to derive a mixed current-density continuity equation for the two fields.
In Sec.~\ref{sec:symmetries} we consider the symmetry of the Hamiltonian under discrete spatial transformations in combination with time reversal, leading to different generalized conservation laws.
We conclude the work in Sec.~\ref{sec:conclusion}.

\section{Two-field Lagrangian}
\label{sec:correlator_lagrangian}

The aim is to construct a general Lagrangian involving wave fields that obey Schr\"{o}dinger equations of motion and allow at the same time to derive generalized types of continuity equations corresponding to different symmetries of the system.
We start by reviewing the variational derivation of the ordinary continuity equation for the probability density $\rho_\circ(\x,t) = |\Psi(\x,t)|^2$ of a single Schr\"odinger field $\Psi$ from the real Lagrangian density \begin{align}\label{eq:3}
\L_\circ &= \Re\left[\Psi^*(i\partial_t - \H_\circ)\Psi \right] \\
	&= \PsiI (\partial_t \PsiR - \H_\circ \PsiI) - \PsiR (\partial_t \PsiI + \H_\circ \PsiR )
\end{align}
where $\PsiR (\PsiI) \in \mathbb{R}$ is the real (imaginary) part of $\Psi = \PsiR + i \PsiI$.
The spatial representation of the Hamilton operator is $\H_\circ = -\frac{1}{2}\nabla^2 + V$ with the real potential function $V(\x)$. 

The Lagrangian is made real by construction to enable unambiguous application of the extremal action principle, treating $\PsiR,\PsiI$ as independent variables. 
Imposing $\delta S_\circ$ for the action $S_\circ = \int_{\Omega} \int_{t_i}^{t_f} d\x~dt~\L_\circ$ under variation of the $\PsiR,\PsiI$, with vanishing variations $\delta \PsiR, \delta \PsiI$ at the boundary of $\Omega$ for any time $t$ and at $t = t_i,t_f$ for all $\x \in \Omega$, produces the Euler-Lagrange equations $\partial_t \PsiR = \H_\circ \PsiI$ and $-\partial_t \PsiI = \H_\circ \PsiR$, which are combined into the Schr\"odinger equation (SE) $i\partial_t \Psi = \H_\circ \Psi$.
The invariance of the $\L_\circ$ under the global $U(1)$ phase transformation $\tilde{\Psi} = e^{i \phi} \Psi$  ($\phi \in \mathbb{R}$) can be used to derive the continuity equation $\partial_t \rho_\circ = - \bs{\nabla} \cdot \J_{\mkern-4mu\circ}$, with the usual current density $\J_{\mkern-4mu\circ} = (\Psi^* \bs{\nabla} \Psi - \Psi \bs{\nabla} \Psi^*)/2i$, by Noether's first theorem.
Throughout, $\bs{\nabla} = \sum_{d = 1,2,3} \hat{\bs{x}}_d \partial/\partial x_d$ denotes differentiation with respect to the coordinate $\bs{x} = \sum_d x_d \hat{\bs{x}}_d $.

This common approach clearly cannot be used to derive continuity equations for nonlocal currents as the ones used in Ref.\,\cite{Kalozoumis2014_PRL_113_050403_InvariantsBroken}. 
It also fails to generate the equations of motion for non-Hermitian Hamiltonians, widely used in effective descriptions. 
Interestingly, both issues can be treated with a single modification of the Lagrangian, by generalizing it to a form which involves two different fields $\Psi_\pm = \PsiR_\pm + i \PsiI_\pm$,
\begin{align}\label{eq:8}
\L &= \Re \left[ \Psi_-^* (i\partial_t - \H)\Psi_+ \right] \\
    &= \PsiI_- \left(\partial_t \PsiR_+ - \H_\circ \PsiI_+ - W \PsiR_+\right) \nonumber  \\
    &- \PsiR_- \left(\partial_t \PsiI_+ + \H_\circ \PsiR_+ - W \PsiI_+\right)
\end{align}
where the Hamiltonian $\H = \H_\circ + i W$ generally includes an additional imaginary term $iW(\x)$ representing a simple model for spatially dependent gain ($W>0$) or loss ($W<0$) of density.

Now, imposing $\delta S = \int_{\Omega}\int_{t_i}^{t_f} d\x dt \delta \L = 0$ under variation of the four independent variables $\PsiR_\pm,\PsiI_\pm$ leads to the Euler-Lagrange equations
\begin{equation}\label{eq:10}
\partial_t \PsiR_\pm = \H_\circ \PsiI_\pm \pm W \PsiR_\pm, ~~~ \partial_t \PsiI_\pm = - \H_\circ \PsiR_\pm \pm W \PsiI_\pm.
\end{equation}
Recombining real and imaginary parts, those in turn yield the two complementary SEs
\begin{equation}\label{eq:SE_dual}
i \partial_t \Psi_\pm = \H_\circ \Psi_\pm \pm i W \Psi_\pm,
\end{equation}
with Hamiltonians $\H_\pm \equiv \H_\circ \pm i W$ for the two fields $\Psi_\pm$ which thus evolve under opposite gain/loss rate profile $W(\x)$.
In this sense, the Lagrangian $\L$ correlates the two states $\Psi_\pm$ via the non-Hermitian part of the effective Hamiltonian.

\section{Generalized mixed continuity equation}
\label{sec:mixed_cont_eq}

We shall now use the introduced two-field Lagrangian to generate a corresponding continuity equation which mixes the two states $\Psi_\pm$.
$\L$ is invariant under the transformation
\begin{equation}\label{eq:13}
\tilde{\Psi}_\pm = e^{\pm \phi} \Psi_\pm,  ~~~~~ \phi = \phiR + i \phiI
\end{equation}
with the real variables $\phiR$ and $\phiI$ parametrizing a dilatation and a rotation in the complex plane, respectively, 
\begin{equation}\label{eq:14}
\left( \begin{array}{c} \tilde{\Psi}^\text{r}_\pm \\ \tilde{\Psi}^\text{i}_\pm \end{array} \right)  = 
e^{\pm \phiR} 
\left( \begin{array}{cc} \cos \phiI & -\sin \phiI \\ \sin \phiI & \cos \phiI \end{array}  \right) \left( \begin{array}{c} \PsiR_\pm \\ \PsiI_\pm  \end{array} \right).
\end{equation}
We can now exploit the invariance of $\L$ under the above transformation to derive conservation laws for the $\PsiR_\pm,\PsiI_\pm$ via Noether's theorem.
To first order in infinitesimal variations $\delta \phiR$ and $\delta \phiI$, the field component variations are 
$\delta_\phi \PsiR_\pm = \pm \PsiR_\pm \delta \phiR - \PsiI_\pm \delta \phiI$ and 
$\delta_\phi \PsiI_\pm = \pm \PsiI_\pm \delta \phiR + \PsiR_\pm \delta \phiI$, leading to a variation
\begin{align}\label{eq:19}
\delta_\phi \L =  
\delta \phiR [ \PsiR_- (\partial_t - W) \PsiI_+ &- \PsiI_- (\partial_t - W) \PsiR_+ \\ 
&+ \PsiR_- \H_\circ \PsiR_+ + \PsiI_- \H_\circ \PsiI_+ ] ~~~~~~~~~~~ \nonumber \\
+~ \delta \phiI [\PsiR_- (\partial_t - W) \PsiR_+ &+ \PsiI_- (\partial_t - W) \PsiI_+ \nonumber \\ 
&- \PsiR_- \H_\circ \PsiI_+ + \PsiI_- \H_\circ \PsiR_+] \nonumber
\end{align}
in the Lagrangian.
Solving the equations of motion (\ref{eq:10}) for the terms $W \PsiR_-$, $W \PsiI_-$ and substituting them in  Eq.~(\ref{eq:19}), the condition $\delta_\phi \L = 0$ yields the following two equations: 
\begin{align}\label{eq:21}
&\partial_t (\PsiR_- \PsiR_+ + \PsiI_- \PsiI_+) = \\ 
&- \frac{1}{2} \bs{\nabla} \cdot [\PsiR_+ \bs{\nabla} \PsiI_- -\PsiI_- \bs{\nabla} \PsiR_+ + \PsiR_-\bs{\nabla} \PsiI_+ - \PsiI_+ \bs{\nabla} \PsiR_-], \nonumber 
\end{align}
resulting from the invariance under phase transformation, and
\begin{align}\label{eq:22}
&\partial_t (\PsiR_- \PsiI_+ - \PsiI_- \PsiR_+) = \\ 
&- \frac{1}{2} \bs{\nabla} \cdot [\PsiI_+ \bs{\nabla} \PsiI_- - \PsiI_- \bs{\nabla} \PsiI_+  + \PsiR_-\bs{\nabla} \PsiR_+ - \PsiR_+ \bs{\nabla} \PsiR_-], \nonumber 
\end{align}
resulting from the invariance under dilatations.
Those equations combine into a single continuity equation 
\begin{equation}\label{eq:mixed_continuity}
\partial_t \rho(\x,t) + \bs{\nabla} \cdot \J(\x,t) = 0  
\end{equation}
for the mixed two-state density and current
\begin{equation}
\rho = \Psi_-^* \Psi_+ , ~~~~ \J = \frac{1}{2i} \left(\Psi_-^* \bs{\nabla} \Psi_+ - \Psi_+ \bs{\nabla} \Psi_-^* \right).  
\end{equation}
As pointed out by Gottfried \cite{Gottfried2003____QuantumMechanics}, such a conservation law stems from the unitary evolution of any two solutions of the SE, rendering their scalar product constant in time.
We here derive this conservation law from the variational principle on the two-field Lagrangian $\L$, and generalize it to non-Hermitian Hamiltonians with $W \neq 0$.
If the net flux of $\J$ at the surface of a domain $\Omega$ vanishes, like when the $\Psi_\pm$ obey von Neumann or Dirichlet boundary conditions on $\Omega$, then Eq.\,(\ref{eq:mixed_continuity}) leads to time-conserved charge $C = \int_\Omega \rho(\x,t) d^D x$ in $D$ dimensions.

%%%%%%%%%%%%%%%%%%%%%%%%%%%%%%%%%%%%%%%%%%%%%%%%% TABLE %%%%%%%%%%%%%%%%%%%%%%%%%%%%%%%%%%%%%%%%%%%%%%%%%%%%%%%%%%%%%%%%%%%%%%%%%%%%%%
\begin{table*} 
  \caption{Generalized conservation laws for the currents 
  \textbf{(a)} $\J_{\mkern-4mu\T}$ in absence of $\F$-symmetry, 
  \textbf{(b)} $\J_{\mkern-4mu\F}$ in presence of $\F$-symmetry, and 
  \textbf{(c)} $\J_{\mkern-4mu\F\T}$ in presence of $\F\T$-symmetry, 
  produced by the mixed continuity equation (\ref{eq:mixed_continuity}) upon substitution of $\Psi_-$ in terms of $\Psi_+$ (subscript then omitted) for non-Hermitian Hamiltonian $\H = \H_\circ + iW$.
  The identity transformation $\F = \mathcal{I}$ reproduces the ordinary probability current conservation in (c).\\
  }
  \bgroup
\def\arraystretch{2}%  1 is default
  \begin{tabularx}{\textwidth}{lccccc} \hline
  \hspace{5em} & {symmetry} & \hspace{5em} & {density $\rho(\x,t)$} & \hspace{5em} & {current $\J(\x,t) \times 2i$}  \\ \hline
    (a)
    &  $\F\H\F^{-1} \neq \H$ 
    & & $\Psi(\x,-t) \Psi(\x,t)$ 
    & & $\Psi(\x,-t) \bs{\nabla} \Psi(\x,t) - \Psi(\x,t) \bs{\nabla} \Psi(\x,-t) $ \\ 
    %%%%%%%%%%%%%%%%%%%%%%%%%%%%%%%%%%%%%%%%%%%%%%%%
    (b)
    &  $\F\H\F^{-1} = \H$  
    & & $\Psi(\F\x,-t) \Psi(\x,t)$  
    & & $\Psi(\F\x,-t) \bs{\nabla} \Psi(\x,t) - \Psi(\x,t) \bs{\nabla} \Psi(\F\x,-t) $ \\  
    %%%%%%%%%%%%%%%%%%%%%%%%%%%%%%%%%%%%%%%%%%%%%%%%
    (c)
    &  $\F\H\F^{-1} = \H^*$  
    & & $\Psi^*(\F\x,t) \Psi(\x,t)$  
    & & $\Psi^*(\F\x,t) \bs{\nabla} \Psi(\x,t) - \Psi(\x,t) \bs{\nabla} \Psi^*(\F\x,t) $ \\ \hline 
    %%%%%%%%%%%%%%%%%%%%%%%%%%%%%%%%%%%%%%%%%%%%%%%%
  \end{tabularx}
  \egroup
\label{tab:generalized_continuity}
\end{table*}
%%%%%%%%%%%%%%%%%%%%%%%%%%%%%%%%%%%%%%%%%%%%%%%%% TABLE %%%%%%%%%%%%%%%%%%%%%%%%%%%%%%%%%%%%%%%%%%%%%%%%%%%%%%%%%%%%%%%%%%%%%%%%%%%%%%

% \newpage
\section{Symmetries and generalized \\ conservation laws}
\label{sec:symmetries}

With the mixed continuity equation (\ref{eq:mixed_continuity}) holding generally for the two arbitrary solutions $\Psi_\pm$ to the SE with $\pm i W$, derived by the \textit{continuous} symmetry of $\L$ in $\phi$, we now use it to derive a set continuity equations for a single state. 
Those are induced by symmetry of the Hamiltonian under a given \textit{discrete} spatial transformation
\begin{equation}
 \F:~\x \to \mathbf{y} = \F(\x) \equiv \F\x,
\end{equation}
potentially combined with the operation of time reversal $\T: t \to -t$, which is here additionally expressed by simple complex conjugation, $\T: i \to -i$.
We consider transformations $\F$ that leave the Laplacian $\nabla^2$ invariant, that is, reflections, translations, and rotations (or combinations thereof) in dimension $D$, so that the $\F$ symmetry of $\H$ is determined by the potential functions $V(\x)$ and $W(\x)$.

For a Hermitian Hamiltonian $\H = \H_\circ = \H^\dagger$, the two SEs (\ref{eq:SE_dual}) become identical ($W = 0$).
Setting $\Psi_- = \Psi_+ \equiv \Psi$ then reproduces the ordinary continuity equation for the probability density $\rho_\circ$ and current $\J_{\mkern-4mu\circ}$ (note that $-i\bs{\nabla}$ is replaced by the kinetic momentum in presence of a vector potential).
If $\H$ is additionally $\F$-symmetric, $\F\H\F^{-1} = \H$, then also $\Psi_-(\x,t) = \Psi_+(\F\x,t) \equiv \Psi(\F\x,t)$ is a solution to the common SE, leading through Eq.\,(\ref{eq:mixed_continuity}) to the conservation of a \textit{bilocal} (or two-point) current 
\begin{equation}
 \J_{\mkern-4mu\F}(\x,t) = \frac{1}{2i} [\Psi^*(\F\x,t) \bs{\nabla} \Psi(\x,t) - \Psi(\x,t) \bs{\nabla} \Psi^*(\F\x,t) ]
\end{equation}
obeying the symmetry-induced continuity equation
\begin{equation}\label{eq:bilocal_continuity}
\partial_t \rho_\F(\x,t) + \bs{\nabla} \cdot \J_{\mkern-4mu\F}(\x,t) = 0  
\end{equation}
with corresponding density $\rho_\F(\x,t) = \Psi^*(\F\x,t)\Psi(\x,t)$.
Notably, such a bilocal picture has been used in an alternative interpretation of the double slit experiment \cite{Jr2015_JPAMT_48_155304_BilocalPicture}.
This symmetry-induced conservation law carries over to the case of a non-Hermitian $\H$ if $W$ in Eq.\,(\ref{eq:SE_dual}) is $\F$-antisymmetric, $W(\F\x) = - W(\x)$, that is, if $\H$ is $\F\T$-symmetric.
The most prominently studied case is that of $\P\T$-symmetric systems mentioned in the introduction.
The associated continuity equation \cite{Bagchi2001_MPLA_16_2047_GeneralizedContinuity} is thereby recovered here setting $\F = \P$.

The complementary SEs also produce an alternative conservation law with respect to time reversal for arbitrary $\H$:
From Eq.\,(\ref{eq:SE_dual}) the time-reversed state $\Psi_-^*(\x,-t) \equiv \Psi(\x,t)$ solves the SE with $+iW$, so that substituting $\Psi_-$ into Eq.\,(\ref{eq:mixed_continuity}) produces the continuity equation
\begin{equation}\label{eq:bitemporal_continuity}
\partial_t \rho_\T(\x,t) + \bs{\nabla} \cdot \J_{\mkern-4mu\T}(\x,t) = 0  
\end{equation}
for the \textit{bitemporal} (or two-time) current 
\begin{equation}
 \J_{\mkern-4mu\T}(\x,t) = \frac{1}{2i}[\Psi(\x,-t)\bs{\nabla}\Psi(\x,t) - \Psi(\x,t)\bs{\nabla}\Psi(\x,-t)]
\end{equation}
corresponding to the generalized density $\rho_\T(\x,t) = \Psi(\x,-t)\Psi(\x,t)$.
Combining this with $\F$-symmetry yields a conservation law for the combined bilocal and bitemporal current
$$ \J_{\mkern-4mu\F\T}(\x,t) = \frac{1}{2i}[\Psi(\F\x,-t)\bs{\nabla}\Psi(\x,t) - \Psi(\x,t)\bs{\nabla}\Psi(\F\x,-t)]$$ 
now obeying the continuity equation
\begin{equation}\label{eq:FT_continuity}
\partial_t \rho_{\F\T}(\x,t) + \bs{\nabla} \cdot \J_{\mkern-4mu\F\T}(\x,t) = 0  
\end{equation}
with density $\rho_{\F\T}(\x,t) = \Psi(\F\x,-t)\Psi(\x,t)$ for $\F$-symmetric $\H$ (with $W(\F\x) = W(\x)$ in the non-Hermitian case).
For a stationary state $\Psi(\x,t) = \Psi(\x)e^{-iEt}$ with energy $E$, the spatial part of $\J_{\mkern-4mu\F}$ and $\J_{\mkern-4mu\F\T}$ reproduce the domainwise spatial invariants of Ref.\,\cite{Kalozoumis2014_PRL_113_050403_InvariantsBroken}.
The continuity equations for the above bilocal and bitemporal currents can be interpreted as a self-correlation of a single particle field at different locations and/or times, resulting from the two-field `correlator' Lagrangian $\L$.
The above generalized conservation laws are listed in Table \ref{tab:generalized_continuity}.

Note here that, in the general case of a non-Hermitian Hamiltonian, its (mixed) expectation value $\bar{H} = \int \Psi_-^* H \Psi_+ d^Dx$ is not conserved, in spite of the functional form of $\L$ not depending explicitly on the time variable $t$.
This can be interpreted by the indirect presence of time in the description via the in- or outflow (`source' or `sink', respectively) of current density in spatial regions with $W(\x) \neq 0$, which changes temporally in general, so that the variational procedure for continuous time translations is not expected to produce a conserved associated Noether charge $\bar{H}$.
Indeed, due to the non-unitary evolution in a general non-Hermitian system, the expectation value of any operator will generally change in time, in accordance with a generalized Heisenberg equation of motion \cite{Baker1990_PRA_42_10_NonHermitianQuantumDynamics}.
Here, the $\Psi_{\pm}$ will generally have components of exponentially increasing or decreasing magnitude in time from the imaginary part of their complex energy eigenvalues. 
For instance, exponential decay is commonly encountered in resonance theory \cite{Moiseyev2011____Non-hermitianQuantum}, and exponential increase in a $\P\T$-symmetric system is discussed in \cite{Zheng2010_PRA_82_010103_MathcalptOptical}, in terms of biorthogonal sets of (complex) energy eigenstates.
In the special case of the spatial transformation $\F = \P$, an alternative (so-called $\mathcal{CPT}$) inner product \cite{Bender2002_PRL_89_270401_ComplexExtension} can be defined, by which the characteristics of unitary evolution are retained \cite{Bender2005_CP_46_277_IntroductionPt-symmetric,Mostafazadeh2003_JPAMG_36_7081_ExactPt} for $\P\T$-symmetric potentials.

Finally we point out that, although derived here for a single-particle Hamiltonian, the bilocal continuity equations above apply equally to many-particle systems with $\F$-symmetric interaction between particles at positions $\{\x_n\}$, with $\bs{\nabla}$ in Eq.\,(\ref{eq:mixed_continuity}) replaced by $\bs{\nabla}_{\mkern-4mu n}$ for the $n$-th particle and with the divergence of the resulting $\J_{\mkern-4mu n}$ summed over $n$; see also Ref.\,\cite{Gottfried2003____QuantumMechanics}.
Typical distance-dependent interaction potentials in isolated systems indeed remain invariant under the global distance-preserving transformations $\F$ considered.
The above variational framework thus unifies the extraction of generalized conservation laws for systems with discrete spatial symmetries.

\section{Conclusions}
\label{sec:conclusion}

We introduce a Lagrangian involving two wave fields which depend on the same $D+1$ spatio-temporal coordinates and solve the Schr\"odinger equation with imaginary potentials of opposite signs. 
The Lagrangian interconnects the variations in time and space of a single fields depending on its invariance under discrete spatial $\F$ symmetries combined with time reversal $\T$, thereby acting as a spatiotemporal correlator.
Specifically, the phase and dilatation invariance of the two-field Lagrangian is used to derive a mixed continuity equation for the two fields.
This is in turn used to generate conservation laws for the generalized current in a single state, in dependence of the Hamiltonian symmetry:
a bitemporal current in absence of symmetry, a bilocal current for $\F$-symmetric Hamiltonians, and the combination of these for $\F\T$ symmetry.
The Hermitian case recovers the ordinary probability density current continuity, and for stationary states the domainwise constant bilocal currents of Ref.\,\cite{Kalozoumis2014_PRL_113_050403_InvariantsBroken} for $D=1$ dimension are reproduced which were also recently measured experimentally \cite{Kalozoumis2015_PRB_92_014303_InvariantCurrents}.
Our approach reveals the origin of symmetry-induced conservation laws in a variational framework and opens up the perspective to construct correlator-Lagrangians as a tool to treat more general field theoretical models.
\\

\section*{Acknowledgements}

We thank J. Schirmer and J. Stockhofe for useful comments on the manuscript. 
P. S. gratefully acknowledges illuminating discussions with M. Bartelmann. 
P. A. K. acknowledges financial support from the IKY Fellowships of Excellence for Postdoctoral Research in Greece - Siemens Program. 
\\

%%%%%%%%%%%%%%% REFERENCES %%%%%%%%%%%%%%%%%%%%%%%%%%%%%%%%%%%%%%%%%%

%

%%%%%%%%%%%%%%% REFERENCES %%%%%%%%%%%%%%%%%%%%%%%%%%%%%%%%%%%%%%%%%%

\end{document}